\documentstyle[aps]{revtex}
\def\tr{\mathop{\rm tr}}
\begin{document}
\title{Stationary axisymmetric SU(2) Einstein-Yang-Mills fields with restricted circularity conditions are Abelian}
\draft
\author{F.J. Chinea and F. Navarro-L\'erida }
\address{Dept. de F\'{\i}sica Te\'orica II,  Ciencias F\'{\i}sicas, \\
Universidad Complutense de Madrid \\
E-28040 Madrid, Spain}
\maketitle
\begin{abstract}
In this paper we prove that in a stationary axisymmetric SU(2) Einstein-Yang-Mills theory the most reasonable circularity conditions that can be considered for the Yang-Mills fields imply in fact that the field is of embedded Abelian type, or else that the metric is not asymptotically flat.
\end{abstract}



Since the discovery of the existence of regular solutions in the SU(2) Einstein-Yang-Mills (EYM) theory \cite{BK} a lot of effort trying to find new interesting non-Abelian solutions has been made and new solutions (most of them in a numerical form) have been found in the presence of symmetries (spherical and axisymmetric cases) and for static and stationary spacetimes (see references in \cite{Volkov}). In this paper we will concentrate on the stationary axisymmetric SU(2) EYM theory. When working with stationary axisymmetric Yang-Mills (YM) fields, most of the (numerical) known solutions make use of the same ansatz: the one introduced by Manton \cite{Manton} and Rebbi and Rossi \cite{Rebbi}. However, this ansatz prevents the use of Weyl's coordinates for the spacetime, as opposed to what can be done in the Einstein-Maxwell theory. The possibility of using Weyl's coordinates would simplify considerably the Einstein equations but the ans\"atze considered up to now in this direction have (numerically) been shown to be incompatible with non-Abelianity and asymptotic flatness. Nevertheless, there seems to be no general argument against the possibility of the existence of an ansatz suitable to merge asymptotic flatness and non-Abelian nature with the use of Weyl's coordinates for the metric. Here, we clarify this point by proving rigorously that a rather natural condition (see equation (\ref{ec10}) below) for fulfilling that ansatz cannot in fact be imposed, as a non-Abelian solution for the stationary axisymmetric SU(2) EYM equations with the appropriate asymptotically flat behavior cannot exist. \par
 \indent In order to show this result, let $\mbox{\boldmath $\xi$}$ and $\mbox{\boldmath $\eta$}$ be the Killing vectors that generate the Abelian group $G_2$ of isometries of the stationary axially symmetric spacetime, $\mbox{\boldmath $\xi$}$ being a timelike vector field and $\mbox{\boldmath $\eta$}$ being a spacelike one with compact periodic trajectories. Owing to the fact that both of them commute, we are able to choose adapted coordinates, say $t$ and $\phi$, such that $\mbox{\boldmath $\xi$} = \partial_t$ and $\mbox{\boldmath $\eta$} = \partial_{\phi}$. We will also assume that the elementary flatness condition for $\mbox{\boldmath $\eta$}$ is satisfied so that the axis is a regular two-dimensional submanifold of the spacetime. We further impose on the stationary axisymmetric spacetime that it admits 2-spaces orthogonal to the group orbits, that is to say, that 
\begin{equation}
\xi \wedge \eta \wedge d\xi = \xi \wedge \eta \wedge d\eta = 0 \label{ec1}   
\end{equation}
holds, where $\xi$ (respectively, $\eta$) is the 1-form corresponding to the vector field $\mbox{\boldmath $\xi$}$ (resp. $\mbox{\boldmath $\eta$}$). Now, if one assumes that the metric contains the axis (or, at least, one of its points), the (Ricci-) circularity theorem \cite{Kundt,Carter} states that (\ref{ec1}) is equivalent to
\begin{equation}
\xi \wedge \eta \wedge R(\mbox{\boldmath $\xi$}) = \xi \wedge \eta \wedge R(\mbox{\boldmath $\eta$}) = 0, \label{ec2}
\end{equation}
where $R(\mbox{\boldmath $v$}) \equiv R_{\mu \nu} v^{\mu} dx^{\nu}$, $R_{\mu \nu}$ being the Ricci tensor. With these assumptions the metric can be written in the Lewis-Papapetrou form \cite{Lewis,Papapetrou}: 
\begin{equation}
ds^2 = -f (dt - \omega d\phi)^2+ f^{-1} [e^{2 \gamma} (d\rho^2 + dz^2) +W^2 d\phi^2], \label{ec3}
\end{equation}
where $f$, $\omega$, $\gamma$, and $W$ are functions of the $\rho$ and $z$ coordinates. \par


\indent Let us now suppose that the matter content is given by non-Abelian SU(2) gauge fields coupled to gravity. We will use the following expression for the YM field $F$:
\begin{equation}
F=dA + A \wedge A, \label{def_F}
\end{equation}
where the YM potential $A$ is an su(2)-valued 1-form, which satisfies
\begin{equation}
A^{\dagger} = -A, \mbox{\quad} \tr{A} =0. \label{rep_A}
\end{equation}
In components:
\begin{equation}
F_{\mu \nu} = A_{\nu , \mu} - A_{\mu , \nu} + [A_{\mu},A_{\nu}], \label{F_components}
\end{equation}
where a comma followed by an index denotes a partial derivative with respect to the corresponding coordinate. The EYM equations read
\begin{eqnarray}
\;&G_{\mu \nu} = 8 \pi G T_{\mu \nu},&\; \label{ec5_a} \\
\;&D_{\mu} F^{\mu \nu} = 0,&\; \label{ec5_b}
\end{eqnarray}
where $G_{\mu \nu}$ is the Einstein tensor; here and in the following we take $c=1$. The YM equations (\ref{ec5_b}) can also be conveniently expressed as
\begin{equation}
d  {}^* F + A \wedge  {}^* F -  {}^* F \wedge A = 0, \label{YM_Javier}
\end{equation}
where $ {}^* F$ denotes the Hodge dual of $F$. The energy-momentum tensor, $T_{\mu \nu}$, and the gauge-covariant derivative, $D_{\mu}$, are defined as
\begin{eqnarray}
\;&\displaystyle T_{\mu \nu} \equiv \frac{1}{2 \pi} \tr \{-F_{\mu \sigma} {F_{\nu}}^{\sigma} + \frac{1}{4} g_{\mu \nu} F_{\alpha \beta} F^{\alpha \beta}\},&\; \label{ec6_a} \\
\;&D_{\mu} \equiv \nabla_{\mu} + [A_{\mu}, \cdot],&\; \label{ec6_b}
\end{eqnarray}
where $\nabla_{\mu}$ denotes the covariant derivative. For a review of EYM fields, see \cite{Volkov}.


\indent By using the Einstein equations (\ref{ec5_a}), conditions (\ref{ec2}) can be rewritten as
\begin{equation}
\xi \wedge \eta \wedge T( \mbox{\boldmath $\xi$}) = \xi \wedge \eta \wedge T(\mbox{\boldmath $\eta$}) = 0, \label{ec7}
\end{equation}
which is just the same as asking $T_{t \rho}$, $T_{t z}$, $T_{\phi \rho}$, and $T_{\phi z}$ to vanish. Following Heusler and Straumann \cite{Straumann} and Heusler \cite{Heusler}, one can write the Ricci-circularity conditions in a more compact form:
\begin{eqnarray}
\tr \{F(\mbox{\boldmath $\xi$}, \mbox{\boldmath $\eta$}) B_{\mbox{\boldmath ${\scriptstyle \xi}$}} + {}^* F(\mbox{\boldmath $\xi$}, \mbox{\boldmath $\eta$}) E_{\mbox{\boldmath ${\scriptstyle \xi}$}}\} &=& 0, \label{ec8_a} \\
\tr \{F(\mbox{\boldmath $\xi$}, \mbox{\boldmath $\eta$}) B_{\mbox{\boldmath ${\scriptstyle \eta}$}} + {}^* F(\mbox{\boldmath $\xi$}, \mbox{\boldmath $\eta$}) E_{\mbox{\boldmath ${\scriptstyle \eta}$}}\} &=& 0, \label{ec8_b}
\end{eqnarray}
where $E_{\mbox{\boldmath ${\scriptstyle v}$}}$ and $B_{\mbox{\boldmath ${\scriptstyle v}$}}$ stand for $-i_{\mbox{\boldmath ${\scriptstyle v}$}} F$ and $i_{\mbox{\boldmath ${\scriptstyle v}$}} {}^* F$, respectively, $i_{\mbox{\boldmath ${\scriptstyle v}$}}$ being the inner product (see \cite{Heusler} for definitions). These conditions are valid for SU(N) in general. \par
\indent The next step is to impose symmetries on the YM fields. Following Bergmann and Flaherty \cite{Bergmann} and Forg\`acs and Manton \cite{Forgacs}, and taking into account that $\mbox{\boldmath $\xi$}$ and $\mbox{\boldmath $\eta$}$ commute, it is possible to use part of the gauge freedom in order to write the symmetries on the YM potentials as
\begin{equation} 
{\cal L}_{{\mbox{\boldmath ${\scriptstyle \xi}$}}} A_{\mu}= {\cal L}_{{\mbox{\boldmath ${\scriptstyle \eta}$}}} A_{\mu} = 0, \mbox{  i.e.,  } A_{\mu} = A_{\mu}(\rho, z). \label{ec9}
\end{equation}
Some gauge freedom still remains, allowing us to perform transformations which depend on $\rho$ and $z$ only, if necessary. \par
\indent Looking at equations (\ref{ec8_a}) and (\ref{ec8_b}), which represent four different constraints, one could ask oneself if they might be a consequence
 of the EYM equations plus the symmetry conditions. That is true for the electromagnetic case, because the relations
\begin{equation}
\xi \wedge \eta \wedge F = \xi \wedge \eta \wedge {}^* F = 0 \label{ec10}
\end{equation}
follow from the Maxwell equations and the fact that $A=A(\rho,z)$. However, as claimed in \cite{Straumann}, in a non-Abelian case there are no known general arguments derived from EYM equations and symmetry conditions that establish $F(\mbox{\boldmath $\xi$},\mbox{\boldmath $\eta$})={}^* F(\mbox{\boldmath $\xi$}, \mbox{\boldmath $\eta$})=0$. Nevertheless, these appear to be the most reasonable assumptions one can put forward in order to fulfill equations (\ref{ec8_a}) and (\ref{ec8_b}). It should be noticed that this ansatz for the YM fields is conserved under gauge transformations, as it is imposed on the YM fields instead of on the YM potentials. \par 
\indent In the following, we will assume the restricted circularity conditions $F(\mbox{\boldmath $\xi$},\mbox{\boldmath $ \eta$}) = {}^* F(\mbox{\boldmath $\xi$}, \mbox{\boldmath $\eta$}) = 0$ (in coordinates, $F_{t \phi} = F_{\rho z}=0$) for stationary axisymmetric SU(2) gauge fields, and we will prove that these natural assumptions give rise to embedded Abelian cases or to non-asymptotically flat spacetimes. Using $F_{t \phi} = F_{\rho z}=0$ in (\ref{ec5_a}), it is very easy to see that $W$ has to be harmonic, i.e., 
\begin{equation}
W_{,\rho \rho} + W_{,z z} = 0. \label{ec11}
\end{equation}
For this reason it is possible to perform a coordinate transformation such that $W = \rho$ (Weyl's coordinates), so that the metric may be written as
\begin{equation}
ds^2 = -f (dt - \omega d\phi)^2+ f^{-1} [e^{2 \gamma} (d\rho^2 + dz^2) +\rho^2 d\phi^2]. \label{ec12}
\end{equation} 
\indent In these coordinates it can be shown that the Einstein equations (\ref{ec5_a}) reduce to the following two equations where neither $\gamma$ nor its derivatives appear, plus other equations for $\gamma$, which will not be relevant in what follows:
\begin{eqnarray}
\;&\displaystyle \rho^2 f \nabla^2 f -\rho^2 (\nabla f)^2 + f^4 (\nabla \omega)^2 = -4 G f \tr \{f^2 (\omega F_{t \rho} - F_{\rho \phi})^2 + f^2 (\omega F_{t z} - F_{z \phi})^2 + \rho^2 (F_{t \rho}^2 + F_{t z}^2)\},&\; \label{ec13_a} \\
\;&\displaystyle \nabla \cdot (\rho^{-2} f \nabla \omega) = -8 G \rho^{-2} f \tr \{F_{t \rho} F_{\rho \phi} + F_{t z} F_{z \phi} - \omega (F_{t \rho}^2+F_{t z}^2)\},&\; \label{ec13_b}
\end{eqnarray}
where $\nabla$ represents the 3-dimensional flat-space nabla operator in cylindrical coordinates $(\rho, z, \phi)$. \par
\indent Let us now concentrate on the YM equations and more concretely on the $\rho$ and $z$ components. In order to simplify them we use the constraint $F_{t \phi} = 0$. Due to our choice of a gauge where $A$ depends on $\rho$ and $z$ only and to the fact that the gauge group is SU(2), the above restriction leads to two possibilities: $A_{\phi}=0$ or $A_t = \lambda A_{\phi}$, with $\lambda=\lambda(\rho,z)$ a scalar function. \par
\indent For the first one it is easy to prove that every component of $F$ has to be proportional to $A_t$. We shall show immediately that this leads to an embedded Abelian field (recall that, by definition, an embedded Abelian field is one such that its potential $A$ satisfies $A=\beta T_0$, where $\beta$ is a scalar 1-form and $T_0$ a constant element ($d T_0 = 0$) in the Lie algebra of the gauge field; the YM equations are then equivalent to the Maxwell equations for the potential $\beta$, $d {}^* (d \beta)=0$). The proof proceeds as follows: If all components of $F$ commute with $A_t$, then they are all proportional to a common element in su(2), and $[F_{\mu \nu}, F_{\alpha \beta}]=0$, for all indices $\mu$, $\nu$, $\alpha$, $\beta$. This can be expressed by the statement
\begin{equation}
F = \sigma T, \label{ecu}
\end{equation}
where $\sigma$ is a scalar 2-form and $T$ is a (in general, coordinate-dependent) 0-form with values in the Lie algebra su(2). By imposing the Bianchi identity $d F + A \wedge F - F \wedge A=0$ on $F$, we get
\begin{equation}
0=d \sigma T + \sigma \wedge (d T + [A,T]). \label{Bianchi}
\end{equation}
We deal succesively with two possible cases: either $d T + [A,T]$ is proportional to $T$ or they are independent. In the first case, $d T + [A,T] = \alpha T$, where $\alpha$ is a scalar 1-form. By exterior differentiation of this equation, we get $d \alpha=0$, thus giving locally $\alpha = d h$ for a certain function $h$. Substituting this in (\ref{Bianchi}), we get $0=d \sigma + \sigma \wedge d h$, so that $d (e^h \sigma)=0$. Then, there will exist locally a 1-form $\beta$ such that
\begin{equation} 
\sigma = e^{-h} d \beta. \label{forma_sigma}
\end{equation}
The YM equation (\ref{YM_Javier}) for $F$ yields $0=e^{-h} d ({}^* d \beta) T$, i.e., $d {}^* (d \beta)=0$, thus showing that the physical content of a YM field satisfying (\ref{ecu}) is simply that of a Maxwell field. By defining $\tilde{T} \equiv e^{-h} T$, we get
\begin{eqnarray}
\;&F=d \beta \tilde{T},&\; \label{Ab_tilde1} \\
\;&d \tilde{T} = \tilde{T} \wedge A - A \wedge \tilde{T},&\; \label{Ab_tilde2} \\
\;&d ({}^* d \beta) = 0.&\; \label{Ab_tilde3}
\end{eqnarray}
In the case where $T$ and $d T + [A,T]$ are independent, we get $d \sigma=0$ and $d T + [A,T]=0$. There will exist locally $\beta$ such that $\sigma = d \beta$. The resulting equations are similar to (\ref{Ab_tilde1}-\ref{Ab_tilde3}):
\begin{eqnarray}
\;&F=d \beta T,&\; \label{Ab1} \\
\;&d T = T \wedge A - A \wedge T,&\; \label{Ab2} \\
\;&d ({}^* d \beta) = 0.&\; \label{Ab3}
\end{eqnarray}
We shall now conclude from (\ref{Ab1}-\ref{Ab3}) that the gauge-invariant condition (\ref{ecu}) is equivalent to the standard definition of embedded Abelian fields (similar considerations apply to the formally identical equations (\ref{Ab_tilde1}-\ref{Ab_tilde3})): By defining
\begin{equation}
C \equiv A -\beta T, \label{def_C}
\end{equation}
and substituting $A= \beta T + C$ in (\ref{Ab1}), we conclude
\begin{equation}
d C + C \wedge C = 0, \label{ec_C}
\end{equation}
so that $C$ is pure gauge (locally, $C=S^{-1} d S$, for an SU(2)-valued 0-form $S$). Thus, $A=\beta T + S^{-1} d S$. By substituting this expression for $A$ in (\ref{Ab2}), we get $d(S T S^{-1})=0$,  so that $S T S^{-1} = T_0$, for a certain $T_0$,  with $d T_0=0$. In conclusion, 
\begin{equation}
A=S^{-1}\beta T_0 S + S^{-1} d S, \label{A_Abelian1}
\end{equation}
so that $A$ is just a gauge transform of
\begin{equation}
A_0 =  \beta T_0. \label{A_Abelian2}
\end{equation}
(Please notice that the gauge transformation $S$ may be chosen to be independent of $t$ and $\phi$, thus leaving condition (\ref{ec9}) invariant).


\indent Let us consider more closely the other option:
\begin{equation}
A_t = \lambda A_{\phi}. \label{ec14}
\end{equation}
When one substitutes (\ref{ec14}) into the $\rho$ and $z$ components of (\ref{ec5_b}) the following relations are obtained:
\begin{eqnarray}
(\lambda f \omega + \rho \lambda + f)(\lambda f \omega - \rho \lambda + f) [A_{\phi}, F_{\phi \rho}]&=&0, \label{ec15_a} \\
(\lambda f \omega + \rho \lambda + f)(\lambda f \omega - \rho \lambda + f) [A_{\phi}, F_{\phi z}]&=&0. \label{ec15_b}
\end{eqnarray}
There are three possible choices. If the commutators vanish, we apply the result just proved above, and we obtain again an embedded Abelian solution, as every component of $F$ can be shown to be proportional to $A_{\phi}$. As for the two other possibilities, they essentially reduce to the same one because they are related by means of a reversal of the sense of rotation  ($\lambda \rightarrow -\lambda$, $\omega \rightarrow -\omega$). For that reason, we can choose one of them, our result being valid for the other one, too. As a consequence, the form for the function of proportionality between $A_t$ and $A_{\phi}$ reads:
\begin{equation}
\lambda = \frac{f}{\rho - f \omega}. \label{ec16}
\end{equation}   
\indent There still remain the two other components of the YM equations, namely, the $t$ and $\phi$ components. Using on the previously mentioned equations the relations 
\begin{eqnarray} 
F^{\rho t} + \frac{1}{\lambda} F^{\rho \phi} & = & -\frac{f}{\rho e^{2 \gamma}} \frac{\lambda_{,\rho}}{\lambda} A_{\phi}, \label{ec17_a} \\
F^{z t} + \frac{1}{\lambda} F^{z \phi} & = & -\frac{f}{\rho e^{2 \gamma}} \frac{\lambda_{,z}}{\lambda} A_{\phi}, \label{ec17_b}
\end{eqnarray}
derived from (\ref{ec14}), one obtains the following equation:
\begin{equation}
\left\{ \lambda_{,\rho \rho} + \lambda_{,z z} - \frac{\rho + f \omega}{\rho \lambda} ({\lambda_{,\rho}}^2 + {\lambda_{,z}}^2) \right\} A_{\phi} = 0. \label{ec18}
\end{equation}
As said before, the case with $A_{\phi}=0$ is an embedded Abelian one, so we will concentrate on the other possibility:
\begin{equation}
 \lambda_{,\rho \rho} + \lambda_{,z z} - \frac{\rho + f \omega}{\rho \lambda} ({\lambda_{,\rho}}^2 + {\lambda_{,z}}^2)=0. \label{ec19}
\end{equation}
\indent The second-order derivatives in (\ref{ec19}) may be substituted by using (\ref{ec16}) and a combination of the Einstein field equations (\ref{ec13_a}) and (\ref{ec13_b}) (to be more precise, we use the field equation corresponding to $-(\rho-f \omega)^2 G_{tt} + 2 f (\rho-f \omega) G_{t \phi} - f^2 G_{\phi \phi}$). The resulting equation reads: 
\begin{equation}
\rho f_{,\rho} + f^2 \omega_{,\rho} - f = -4G \frac{f^2}{\lambda^2} ({\lambda_{,\rho}}^2 + {\lambda_{,z}}^2) \tr ({A_{\phi}}^2). \label{ec20}
\end{equation}
From (\ref{ec20}) we see that the left-hand side has to be non-negative (recall that $\tr ({A_{\phi}}^2) < 0$ !), but in that case the asymptotically flat condition cannot hold. To prove that, one only has to introduce in (\ref{ec20}) the asymptotic behavior of $f$ and $\omega$ 
\begin{eqnarray}
f & \longrightarrow & 1 - \frac{2 M}{r} + O\left( \frac{1}{r^2} \right), \label{ec21_a} \\
\omega & \longrightarrow & -\frac{2 J}{r} \sin^2 \theta + O\left( \frac{1}{r^2} \right), \label{ec21_b}
\end{eqnarray}
where $r$ and $\theta$ are spherical coordinates related to $\rho$ and $z$ as $r=\sqrt{\rho^2 + z^2}$ and $\theta=\arctan(\rho/z)$, and $M$ and $J$ are constants. When this is done, the leading term of the left-hand side goes like $-1$, yielding a contradiction. \par
\indent Thus, the only case which is not essentially Abelian has  to be non-asymptotically flat, which makes it unacceptable. We have proved this result for an SU(2) EYM theory. Our method depends on the fact that for SU(2) a vanishing commutator of two quantities in the corresponding Lie algebra implies that either one of them vanishes, or that a relation such as (\ref{ec14}) holds. However, this is not true for SU(N) in general, because if N is greater than two, it is possible to find two-dimensional Abelian subalgebras in the associated Lie algebra. Therefore, the procedure followed here cannot be generalized to SU(N) in general.  \par 
\indent The present work has been supported in part by DGICYT Project PB98-0772; F.N.L. is supported by an FPI Predoctoral Scholarship from Ministerio de Educaci\'on (Spain). The authors wish to thank L. Fern\'andez-Jambrina, L. M. Gonz\'alez-Romero, and M. J. Pareja
for discussions.



%
%

%
%

\end{document}